\documentclass[apj,twocolumn]{oja}

\usepackage[T1]{fontenc}
\usepackage{aas_macros}
\usepackage{natbib}
\usepackage{times}

\usepackage{amsmath}
\usepackage{booktabs}
\usepackage{multirow}
\usepackage{color}
\usepackage{soul}
\usepackage[breaklinks,colorlinks,citecolor=blue,urlcolor=blue]{hyperref}
\usepackage{orcidlink}
\usepackage{xspace}
\usepackage{xurl}

\newcommand{\package}[1]{\textsl{#1}}

\newcommand{\myr}{\mbox{${\rm Myr}$}}
\newcommand{\gyr}{\mbox{${\rm Gyr}$}}
\newcommand{\pc}{\mbox{${\rm pc}$}}

\newcommand{\kpc}{\mbox{${\rm kpc}$}}

\newcommand{\tgmc}{\mbox{$t_{\rm GMC}$}}
\newcommand{\tover}{\mbox{$t_{\rm over}$}}
\newcommand{\rgmc}{\mbox{$R_{\rm GMC}$}}

\newcommand{\hii}{H{\sc ii}\xspace}

\newcommand{\be}{\begin{equation}}
\newcommand{\ee}{\end{equation}}
\newcommand{\bea}{\begin{eqnarray}}
\newcommand{\eea}{\end{eqnarray}}

\defcitealias{koda23}{KT23}

\received{2024 April 24}
\revised{X{\small xxxx} XX}
\accepted{X{\small xxxx} XX}

\shorttitle{Death of the Immortal Molecular Cloud}
\shortauthors{Kruijssen et al.}

\begin{document}\sloppy\sloppypar\raggedbottom\frenchspacing

\title{Death of the Immortal Molecular Cloud: Resolution Dependence of the Gas-Star Formation Relation Rules out Decoupling by Stellar Drift\vspace{-15mm}}

\author{J.~M.~Diederik~Kruijssen$^\star$\,\orcidlink{0000-0002-8804-0212}$^{1,2}$}
\author{M\'{e}lanie~Chevance\,\orcidlink{0000-0002-5635-5180}$^{3,2}$}
\author{Steven~N.~Longmore\,\orcidlink{0000-0001-6353-0170}$^{4,2}$}
\author{Adam Ginsburg\,\orcidlink{}$^{5}$}
\author{\\Lise Ramambason\,\orcidlink{}$^{3}$}
\author{Andrea Romanelli\,\orcidlink{}$^{3}$}
\thanks{$^\star$E-mail: \href{mailto:kruijssen [at] coolresearch.io}{kruijssen [at] coolresearch.io}}

\affiliation{$^1$Technical University of Munich, School of Engineering and Design, Department of Aerospace and Geodesy, Chair of Remote Sensing Technology, Arcisstr.~21, 80333 Munich, Germany}
\affiliation{$^2$Cosmic Origins Of Life (COOL) Research DAO, \href{https://coolresearch.io}{https://coolresearch.io}}
\affiliation{$^3$Institut f\"ur Theoretische Astrophysik, Zentrum f\" ur Astronomie der Universit\"at Heidelberg, Albert-\"Uberle-Str.~2, 69120 Heidelberg, Germany}
\affiliation{$^4$Astrophysics Research Institute, Liverpool John Moores University, IC2, Liverpool Science Park, 146 Brownlow Hill, Liverpool L3 5RF, UK}
\affiliation{$^5$Department of Astronomy, University of Florida, P.O.\ Box 112055, Gainesville, FL, USA\\}


\begin{abstract}\noindent
Recent observations have demonstrated that giant molecular clouds (GMCs) are short-lived entities, surviving for the order of a dynamical time before turning a few percent of their mass into stars and dispersing, leaving behind an isolated young stellar population. The key question has been whether this GMC dispersal actually marks a point of GMC destruction by stellar feedback from the new-born stars, or if GMCs might be `immortal' and only dynamically decouple from their nascent stars due to stellar drift. We address this question in six nearby galaxies, by quantifying how the gas-star formation relation depends on the spatial scale for scales between the GMC diameter and the GMC separation length, i.e.\ the scales where an excess of GMCs would be expected to be found in the stellar drift scenario. Our analysis reveals a consistent dearth of GMCs near young stellar populations regardless of the spatial scale, discounting the notion of `immortal' GMCs that decouple from their nascent stars through stellar drift. Instead, our findings demonstrate that stellar feedback destroys most GMCs at the end of their lifecycle. Employing a variety of statistical techniques to test both hypotheses, we find that the probability that stellar feedback concludes the GMC lifecycle is about 2,000 times higher than the probability that stellar drift separates GMCs and young stellar regions. This observation strengthens the emerging picture that galaxies consist of dynamic building blocks undergoing vigorous, feedback-driven lifecycles that collectively regulate star formation and drive the baryon cycle within galaxies.\\
\end{abstract}

\section{Introduction}
\label{sec:intro}
The vast majority of star formation in galaxies takes place in giant molecular clouds (GMCs; \citealt{chevance23}). The cycle of GMC collapse, star formation, GMC dispersal, and stellar feedback represents the engine driving galaxy evolution. As a result, understanding the GMC lifecycle has been one of the key open questions in modern studies of star formation and galaxy evolution.

Until recently, the key unknown was on what timescales the GMC lifecycle takes place, with qualitative implications for the physical processes driving the matter cycle within galaxies. The traditional picture of long-lived GMCs \citep[e.g.][]{scoville79,koda09}, with lifetimes only a factor of several shorter than the galaxy wide gas depletion time of $\sim1~\gyr$ \citep[e.g.][]{leroy13}, would imply that GMCs turn a large fraction of their mass into stars over their lifetimes, requiring modest mass loading factors and mass outflow rates. By contrast, the dynamical picture of star formation taking place within short-lived GMCs over the course of a crossing time \citep[e.g.][]{elmegreen00,hartmann01} would imply considerably higher mass loading factors and mass outflow rates -- not necessarily driving gas out of the galaxy at large \citep{keller22}, but certainly requiring a violent matter cycle on sub-galactic scales. Distinguishing between these two perspectives requires knowledge of the GMC lifetime, measured from the initial GMC condensation out of the atomic phase to the termination of star formation within the GMC and its eventual dispersal.

Methods for inferring GMC lifetimes have existed for a long time \citep[e.g.][]{scoville79,leisawitz89}, but they often relied on strong assumptions, such as requiring GMCs to form at special locations within the host galaxy and follow evolutionary streamlines \citep{koda09,meidt15,kruijssen15}, or requiring GMCs and \hii regions to be easily identifiable entities that can be separated from their environment \citep{kawamura09,corbelli17}, despite the hierarchical structure of the interstellar medium (ISM) and young stellar populations. Either of these approaches carry a high degree of subjectivity. Similarly worryingly, the estimates of the resulting GMC lifetimes reported by these studies varied by (more than) two orders of magnitude, from $1{-}100~\myr$. The heterogeneity of the techniques for measuring the GMC lifetime made it challenging to assess whether this spread was due to physical variations or differences in methodology.

Obtaining robust measurements of the GMC lifetime has recently been possible thanks to two key developments. First, the spatial resolution and sensitivity required to resolve the molecular gas distribution of galaxies (usually traced by CO) into individual GMCs is now routinely achieved \citep[e.g.][]{leroy21_survey}. Second, statistically robust and objective measurement techniques have now been developed to measure GMC lifetimes. Specifically, the `uncertainty principle for star formation' initially developed by \citet{kruijssen14} and \citet{kruijssen18} (with important extensions by \citealt{hygate19} and \citealt{haydon20_ext,haydon20}) has enabled routine measurements of the GMC lifetime, the gas clearance timescale after the emergence of the first massive stars (also referred to as the `feedback timescale'), the integrated star formation efficiency per GMC, as well as many other derived quantities, across $50{-}100$ galaxies \citep[e.g.][]{kruijssen19,chevance20,chevance22,ward20,ward22,zabel20,kim21,kim22,kim23,lu22}. These measurements have resulted in GMC lifetimes ranging from $5{-}35~\myr$ across the nearby galaxy population. Typical uncertainties are a few 10s of percent, implying that this range indicates physical variation of the GMC lifetime. Regardless of these quantitative galaxy-to-galaxy variations, these studies have painted a picture of low-efficiency star formation in short-lived GMCs, with high mass loading factors and outflow rates driven by strong stellar feedback.

The uncertainty principle methodology is inspired by the discovery of \citet{schruba10} that the tight relation between the gas mass and the star formation rate (SFR) on large ($\sim\kpc$) scales in galaxies \citep[e.g.][]{kennicutt98,bigiel08,leroy08} breaks down on the small ($\sim100~\pc$) scales of GMCs and \hii regions \citep[also see e.g.][]{onodera10}. This `decorrelation' was quantified by \citet{schruba10} in the form of a `tuning fork diagram', in which the gas-to-SFR tracer flux ratio deviates from the galactic average towards small spatial scales. The deviation is upward or downward, depending on whether the measurement apertures are centred on gas emission peaks (e.g.\ GMCs) or SFR tracer peaks (e.g.\ \hii regions), respectively. The magnitude and (a)symmetry of these deviations in the gas-to-SFR tracer flux ratio directly probe the relative rarities of emission peaks in both tracers, and thus can be used to infer the ratios of their lifetimes, as well as the time for which both tracers coexist within these emission peaks. In practice, these quantities are obtained by fitting the size scale dependence of the gas-to-SFR tracer flux ratio with a mathematical expression that depends on the key evolutionary timescales (GMC lifetime and the GMC-\hii region coexistence timescale) and the region separation length. \citet{kruijssen14} and \citet{kruijssen18} formalised this timescale dependence and provided a robust statistical framework for measuring the timescales that define the GMC lifecycle using the `tuning fork diagrams' of \citet{schruba10}.

No other technique for measuring GMC lifetimes has been subjected to the level of scrutiny that the uncertainty principle methodology has received, ranging from simplified numerical experiments \citep[e.g.][]{kruijssen18,hygate19,haydon20_ext,haydon20} to state-of-the-art numerical simulations \citep[e.g.][]{fujimoto19,jeffreson21,semenov21,keller22} and extensive observational tests \citep[e.g.][]{kruijssen19,chevance20,ward22}. The analysis code (named {\sc Heisenberg}) is publicly available at \href{https://github.com/mustang-project/heisenberg}{https://github.com/mustang-project/heisenberg}. As a result, the measurements themselves are robust and have well-defined uncertainties, but their physical interpretation has spurred lively discussion in the literature.

The GMC lifetimes obtained with the uncertainty principle methodology are similar to a dynamical time, indicating rapid star formation, and the short-lived ($1{-}6~\myr$) spatial coincidence between GMCs and unembedded massive stars suggests that GMC lifetimes are curtailed by early, pre-supernova feedback mechanisms \citep[see discussions in e.g.][]{kruijssen19,chevance20,chevance22,kim22}. The important role of feedback in driving GMC dispersal is further supported by the close match between the outflow velocities implied by these short timescales and direct measurements of the expansion velocity of \hii regions \citep[e.g.][]{mcleod19,mcleod20,mcleod21,ward22} and the implied chemical mixing scale in the ionised medium \citep{kreckel20}.

An alternative hypothesis for explaining the decorrelation between GMCs and their nascent stellar populations was recently proposed by \citet[hereafter \citetalias{koda23}]{koda23}. These authors performed a simple numerical experiment to show that the observed decorrelation may also be produced if the GMCs are in fact immortal, i.e.\ they have `infinite' lifetimes (i.e.\ much longer than a star formation cycle), but the young stars experience a kinetic drift relative to their natal GMC. As the physical agent for this drift, the authors propose cloud-cloud collisions. The feasibility of cloud-cloud collisions as an important mechanism for driving star formation in galaxies has been questioned in numerous studies due to the low predicted incidence of cloud-cloud collisions, as well as other fine-tuning issues \citep[e.g.][]{mckee07,dobbs15,chevance20b}. However, regardless of the physical mechanism causing stellar drift, the drift itself is a valid hypothesis for the observed decorrelation between GMCs and \hii regions.

In earlier papers, stellar drift was disregarded as the cause for the small-scale gas-SFR decorrelation, because either the typical drift velocity was considered too slow \citep{kruijssen14}, or the obtained GMC lifetimes were found to be independent of the presence of spiral arms \citep{chevance20}, contrary to the expectation of stellar drift caused by cloud-cloud collisions. However, the renewed interest in the stellar drift hypothesis by \citetalias{koda23} warrants a more rigorous test of this idea. In this paper, we carry out a simple observational experiment using existing data, which enables a statistically robust distinction between GMC destruction by stellar feedback and GMC isolation due to stellar drift. A description of the experiment and the observational data used are provided in \S\ref{sec:method}. In \S\ref{sec:results}, we present the results of the experiment, showing that the decorrelation cannot be driven by stellar drift, and instead GMCs must be destroyed by stellar feedback. We conclude in \S\ref{sec:disc} with a short discussion of the implications of these results.

\section{Method}
\label{sec:method}

\subsection{Experiment design}
\label{sec:exp}

\begin{figure*}
\includegraphics[width=\hsize]{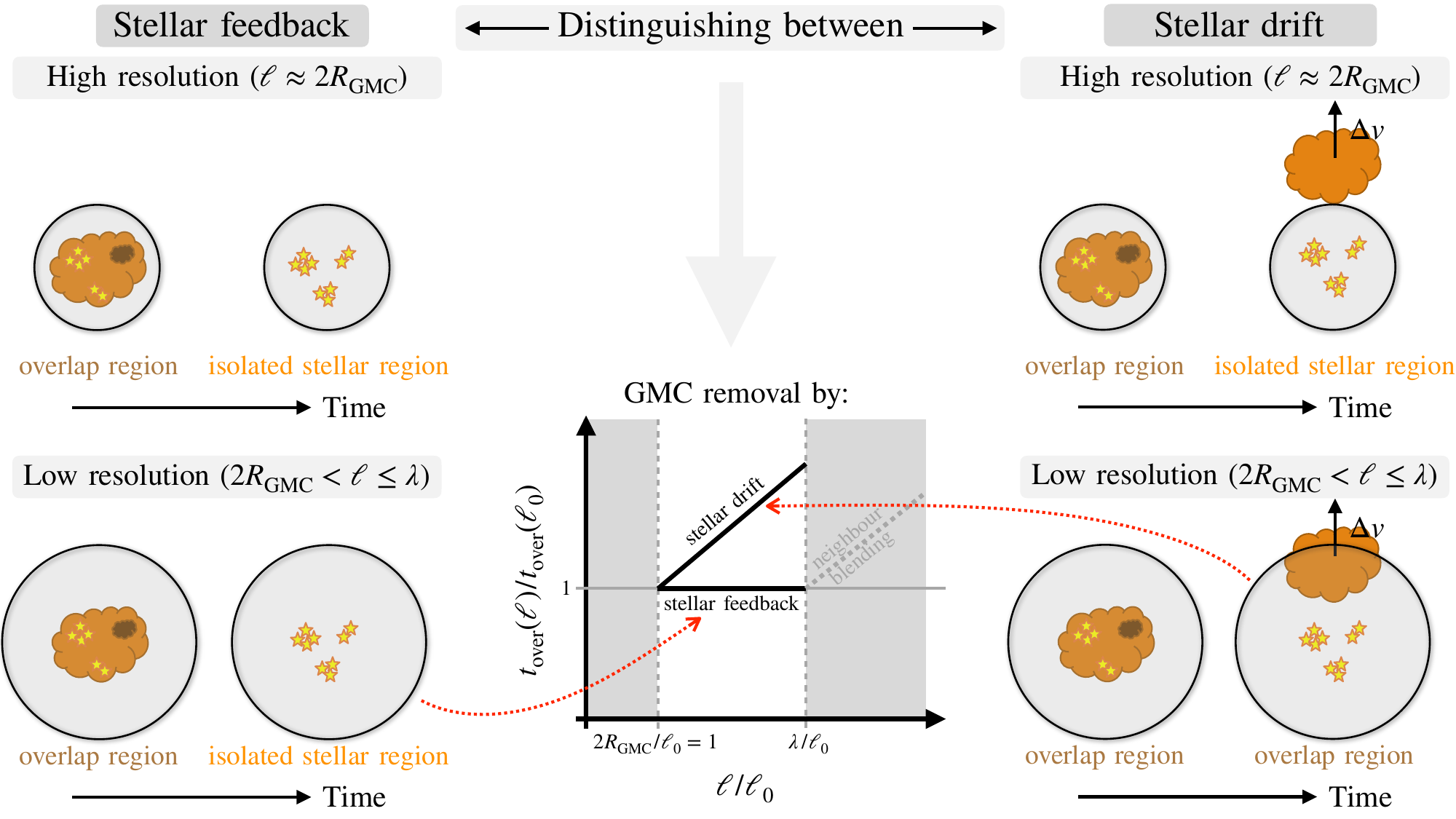}%
\caption{
\label{fig:schematic}
Schematic representation of the experiment performed in this paper. By placing apertures of increasing size on peaks of molecular gas (GMCs) and SFR tracer emission (\hii regions), we measure the coexistence timescale (i.e.\ the `overlap timescale') of these phases using the uncertainty principle methodology \citep{kruijssen18}. We then worsen the resolution at which this measurement is made, from the GMC scale ($\ell\approx2\rgmc$) to the region separation length ($\ell\approx\lambda$). If GMCs and young stellar regions decouple by feedback-driven GMC destruction, then the overlap timescale should be independent of the size scale (i.e.\ aperture size or spatial resolution). If GMCs and young stellar regions decouple by stellar drift, then the overlap timescale should linearly increase with the size scale. On size scales $\ell>\lambda$, the measured overlap timescale may increase regardless of the physics due to blending with neighbouring regions. See \S\ref{sec:exp} for further details.
}
\end{figure*}

We consider a population of GMCs that all form a young stellar population after some time delay. After a further time delay, during which the GMC and stars coincide and which we refer to as the `overlap timescale' or $\tover$, the GMC is removed and the stars reside in isolation. We examine the following two physical scenarios for the removal of the GMC.
\begin{enumerate}
\item
\textit{Stellar feedback}: the GMC is destroyed by stellar feedback, either through a phase transition or by fragmentation into many small parts that observationally give the impression of a diffuse molecular phase. In this case, the GMC lifecycle ends.
\item
\textit{Stellar drift}: the GMC is not destroyed, but is displaced from the stellar population due to kinetic drift, caused by an unspecified physical mechanism (cloud-cloud collisions in \citetalias{koda23}). In this case, the GMC is `immortal' and continues its lifecycle.
\end{enumerate}
Fundamentally, the `GMC lifetime' obtained through the uncertainty principle methodology as described by \citet{kruijssen18} measures the time spent by GMCs prior to and during the appearance of a young stellar population. Likewise, the `overlap timescale' measures the time during which a GMC and its nascent stars are co-spatial. Therefore, the methodology cannot directly distinguish between the two scenarios outlined above.\footnote{The GMC lifecycle measurements obtained so far do directly rule out the existence of immortal GMCs that do \textit{not} dynamically decouple from their nascent stellar populations, because isolated \hii regions exist. In the tuning fork representation, this decoupling manifests itself as a non-zero excess of the gas-to-SFR tracer flux ratio when focusing small apertures on peaks of gas emission. If the young stellar populations were to remain co-spatial with their parent GMC and disappear by stellar evolutionary fading, then the apertures focused on GMCs would always capture all of the SFR tracer flux, and there would be no excess in the gas-to-SFR tracer flux ratio \citep[e.g.][]{fujimoto19}.} However, the dependence of the measured overlap timescale on the size scale fundamentally differs in both cases.
\begin{enumerate}
\item
If stellar feedback destroys GMCs after the stars have formed, then placing increasingly large apertures on isolated stellar populations (i.e.\ post-GMC destruction) would not capture any gas emission from their natal GMCs, because these have either disappeared altogether or have been dispersed into a diffuse phase (which is filtered out in the uncertainty principle methodology and can no longer be identified as a GMC, see \citealt{hygate19}). As a result, \textit{the measured overlap timescale would be independent of the size scale} for aperture sizes between the GMC diameter and the typical separation length between independent GMCs or \hii regions.
\item
If stellar drift causes a spatial displacement between the stellar population and their natal GMCs, then placing increasingly large apertures on isolated stellar populations (i.e.\ post-displacement) would capture the gas emission from their natal GMCs for an increasing fraction of the total number of stellar regions as the aperture size grows. As a result, \textit{the measured overlap timescale would increase with the size scale} for aperture sizes in between the GMC diameter and the typical separation length between independent GMCs or \hii regions.
\end{enumerate}

Both situations are illustrated schematically in \autoref{fig:schematic}. Mathematically, the dependence of the overlap timescale on the size scale can be expressed as follows. For stellar feedback-driven GMC destruction, the overlap timescale measured on a size scale $\ell$ is identical to the overlap timescale measured at the GMC diameter $\ell_{\rm 0}\equiv2\rgmc$, as long as the size scale remains smaller than the region separation length $\lambda$.\footnote{For reference, the region separation length is defined as the mean geometric distance between regions (of any type) in the vicinity of identified emission peaks. As such, it is larger than the nearest-neighbour distance between regions, but smaller than the separation length obtained by uniformly distributing regions across the entire field of view. See \citet{kruijssen14} and \citet{kruijssen18} for a formal definition, and the Methods section of \citet{kruijssen19} for a comparison to other characteristic size scales.} Beyond this size scale, contamination by neighbouring regions artificially increases the overlap timescale (see \S4.3.6 of \citealt{kruijssen18} and fig.~3 of \citealt{kruijssen19}). Therefore, we expect
\begin{equation}
\label{eq:toverfb}
\left[\frac{\tover(\ell)}{\tover(\ell_0)}\right]_{\rm fb}=1~~~~~{\rm for}~~2\rgmc\leq\ell\leq\lambda,
\end{equation}
where the subscript `fb' indicates stellar feedback and $\rgmc$ represents the GMC radius. The GMC radius is obtained as a byproduct of the uncertainty principle methodology and is retrieved directly from the emission structure in the molecular gas maps. This means that if the GMC radius is not resolved, it will be approximately equal to half the resolution scale by definition, i.e.~$\rgmc\approx\ell/2$ (see \S3.2.11 of \citealt{kruijssen18}). Throughout this paper, we define $\ell_0=2\rgmc$, where $\rgmc$ is obtained by performing the uncertainty principle analysis at the best available resolution.

By contrast, the measured overlap timescale in the case of stellar drift is equal to the time it takes a moving stellar population to clear a distance equal to the aperture radius, or half the aperture size $\ell$, at the drift velocity $v_{\rm drift}$.\footnote{For simplicity, we assume a constant drift velocity here, but we will relax this assumption in \S\ref{sec:alpha}.} As before, the requirement is that the size scale is larger than the GMC diameter and remains smaller than the region separation length $\lambda$. In this regime, we physically expect $\tover(\ell)=\ell/2v_{\rm drift}$ and can take the ratio relative to the overlap timescale at the resolution equal to the GMC diameter ($\ell_0$) to obtain the simple expectation
\begin{equation}
\label{eq:toverdrift}
\left[\frac{\tover(\ell)}{\tover(\ell_0)}\right]_{\rm drift}=\frac{\ell}{\ell_0}~~~~~{\rm for}~~2\rgmc\leq\ell\leq\lambda,
\end{equation}
where the subscript `drift' indicates stellar drift.

The qualitatively different behaviour of $\tover(\ell)$ for stellar feedback-driven GMC destruction and stellar drift-driven decoupling means that we can distinguish between both scenarios by measuring the overlap timescale using the uncertainty principle methodology with increasing minimum aperture sizes, in the range $2\rgmc\leq\ell\leq\lambda$, and testing whether the resulting data are more consistent with equation~(\ref{eq:toverfb}) or with equation~(\ref{eq:toverdrift}). As shown in \S\ref{sec:results}, the uncertainties in the measurements are small enough to unambiguously distinguish between both cases.

\begin{table}
\caption{Galaxy sample}
\label{tab:sample}
\centering
\begin{tabular}{lccccc}
\hline
Galaxy & $\ell_{\rm min}$ & $\ell_0$ & $\lambda$ & $\lambda/\ell_{\rm min}$ & H$_2$ tracer   \\
 & [\pc] & [\pc] & [\pc] & & \\
\hline
LMC & $25$ & $24.1^{+0.5}_{-0.3}$ & $71^{+13}_{-8}$ & $2.8^{+0.6}_{-0.3}$ & CO($1{-}0$) \\
NGC300 & $20$ & $30.9^{+2.3}_{-1.6}$ & $113^{+25}_{-17}$ & $5.7^{+1.2}_{-0.9}$ & CO($1{-}0$)  \\
M33 & $49$ & $68.9^{+2.8}_{-3.1}$ & $155^{+30}_{-24}$ & $3.2^{+0.6}_{-0.5}$ & CO($1{-}0$)  \\
NGC628 & $54$ & $52.0^{+2.3}_{-2.9}$ & $96^{+13}_{-10}$ & $1.8^{+0.2}_{-0.2}$ & CO($2{-}1$)  \\
NGC5068 & $35$ & $35.0^{+0.9}_{-0.5}$ & $118^{+19}_{-14}$ & $3.7^{+0.5}_{-0.4}$ & CO($2{-}1$)  \\
M83 & $45$ & $46.6^{+2.1}_{-1.7}$ & $86^{+9}_{-7}$ & $1.9^{+0.2}_{-0.1}$ & CO($2{-}1$)  \\
\hline
\end{tabular}
\end{table}
The overlap timescales are obtained at different size scales by simply omitting the observed gas-to-SFR tracer flux ratios at aperture sizes smaller than $\ell$. This effectively implies truncating the tuning fork diagram before re-fitting the mathematical expression through which the GMC lifetime, overlap timescale, and region separation length are obtained. When carrying out the fit at lower resolutions, we fix the region separation length to the value that was obtained when including all aperture sizes (down to $\ell_0$), to ascertain that the lower-resolution fits are not affected by any degeneracies between the timescales and the region separation length. The resulting overlap timescales are then normalised to the overlap timescale obtained at the best resolution as in equations~(\ref{eq:toverfb}) and~(\ref{eq:toverdrift}).

\subsection{Observational data}
\label{sec:data}

The experiment described in \S\ref{sec:exp} requires observations at a resolution sufficient to probe the gas-to-SFR tracer ratio at scales smaller than the region separation length $\lambda$. All previous applications of the uncertainty principle methodology have yielded measurements of $\lambda$, making it possible to use the ratio $\lambda/\ell_{\rm min}\ga2$, where $\ell_{\rm min}$ is the highest available resolution, as a selection criterion to guarantee a sufficient dynamic range. Additionally, we prioritise absolute spatial resolutions of $\ell_{\rm min}\la50~\pc$ (implying an effective distance cut given the current state-of-the-art in submillimeter observations of the molecular ISM). As the molecular gas tracer, we use CO emission (either the $1{-}0$ or $2{-}1$ transition), and we use H$\alpha$ to trace the SFR. The resulting sample of six galaxies is given in Table~\ref{tab:sample}, listing the best spatial resolution $\ell_{\rm min}$, the estimated GMC diameter $\ell_0$, the region separation length $\lambda$, and the dynamic range ratio $\lambda/\ell_{\rm min}$. The sample contains five galaxies that have been previously analysed -- the LMC by \citet{ward22}, NGC300 by \citet{kruijssen19}, M33 by \citet{kim21}, and NGC628 and NGC5068 by \citet{chevance20}. We refer to these papers for descriptions of the observational data used. In addition, we also include M83, because of its favourable properties (high resolution and sensitivity), and the fact that \citet{koda23b} recently estimated a GMC lifetime of $\sim100~\myr$ in this galaxy by assuming that GMCs form in spiral arms and follow evolutionary streamlines. Because the uncertainty principle methodology does not make such stringent assumptions, it is a worthwhile exercise to also obtain a GMC lifetime measurement for this galaxy. To achieve this, we use the CO($2{-}1$) map from ALMA programmes 2013.1.01161.S, 2015.1.00121.S, and 2016.1.00386.S \citep[PI K.~Sakamoto, processed as described by][]{leroy21_pipe}, and the H$\alpha$ map from SINGG \citep{meurer06}, and perform the uncertainty principle analysis in exactly the same way as in our previous work \citep[e.g.][]{kruijssen19,chevance20,kim22}.

\section{Results}
\label{sec:results}

\subsection{Change of the measured overlap timescale as a function of the spatial resolution}

For each of the six sample galaxies, Table~\ref{tab:fits} lists the GMC lifetime, overlap timescale (in other papers often referred to as the `feedback timescale'), and the region separation length. Galaxies with `updated' in their reference column have had the analysis repeated with the latest version of the analysis pipeline from \citet{kim22}. The changes relative to the originally published numbers fall within the uncertainties. For M83, the table lists new results that are presented for the first time in this paper. We see that the GMC lifetime ($\tgmc=33.2^{+5.6}_{-3.7}~\myr$) and overlap timescale ($\tover=3.9^{+1.1}_{-0.6}~\myr$) fall within the range of previous results ($\tgmc=5{-}35~\myr$ and $\tover=1{-}6~\myr$, respectively, see \citealt{kim22}), indicating that GMC in M83 experience rapid lifecycles, like GMCs in all other nearby galaxies. These results also imply that the small-scale decorrelation of GMCs and \hii regions in M83 is inconsistent with the GMC lifetimes of $\sim100~\myr$ estimated by \citet{koda23b}.\footnote{There are two galaxies for which $\sim100~\myr$ GMC lifetimes have been claimed in the literature. Next to M83, this also includes M51 \citep{koda09}. Interestingly, our GMC lifetime measurements for both of these galaxies reside towards the upper end of the observed range of $\tgmc=5{-}35~\myr$. While for M83 we here obtain $\tgmc=33.2^{+5.6}_{-3.7}~\myr$, the GMC lifetime in M51 is found to be $\tgmc=30.5^{+9.5}_{-4.8}~\myr$ \citep{chevance20}. Consulting the environmental trends of $\tgmc$ identified by \citet{kim22}, these galaxies may fall towards the top of the range because they both have relatively high molecular gas surface densities. In such environments, the observed GMC lifetimes are typically longer.}

\begin{table}
\caption{Uncertainty principle measurement results}
\label{tab:fits}
\centering
\begin{tabular}{lcccc}
\hline
Galaxy & $\tgmc$ [\myr] & $\tover$ [\myr] & $\lambda$ [\pc] & Reference  \\
\hline
LMC & $11.1^{+1.6}_{-1.7}$ & $1.2^{+0.2}_{-0.2}$ & $71^{+13}_{-8}$ & 1 \\
NGC300 & $10.6^{+1.4}_{-1.5}$ & $1.2^{+0.2}_{-0.1}$ & $113^{+25}_{-17}$ & 2, updated  \\
M33 & $14.5^{+1.6}_{-1.5}$ & $3.3^{+0.6}_{-0.5}$ & $155^{+30}_{-24}$ & 1  \\
NGC628 & $24.0^{+2.4}_{-2.4}$ & $3.2^{+0.5}_{-0.6}$ & $96^{+13}_{-10}$ & 3  \\
NGC5068 & $9.0^{+1.5}_{-1.3}$ & $1.3^{+0.3}_{-0.3}$ & $118^{+19}_{-14}$ & 3, updated  \\
M83 & $33.2^{+5.6}_{-3.7}$ & $3.9^{+1.1}_{-0.6}$ & $86^{+9}_{-7}$ & this work  \\
\hline
References: & \multicolumn{4}{l}{(1) \citet{kim21}, (2) \citet{kruijssen19},}\\
& \multicolumn{4}{l}{(3) \citet{kim22}}
\end{tabular}
\end{table}

The key question then is whether the short GMC lifetimes listed in Table~\ref{tab:fits} reflect truly short lifetimes due to GMC destruction by feedback, or whether the GMCs might be `immortal' and decouple from the nascent stellar populations by stellar drift. In the latter case, the short lifetimes would actually represent short periods of time in between bursts of star formation. To answer this question, we perform the experiment outlined in \S\ref{sec:exp} and \autoref{fig:schematic}. For each of the six galaxies, we investigate how the overlap timescale changes if we increase the smallest aperture size down to which we perform the uncertainty principle fit, for aperture sizes in between the GMC diameter ($2\rgmc$) and the region separation length ($\lambda$). If the decoupling between GMCs and \hii regions is caused by stellar drift, then larger apertures focused on \hii regions are more likely to contain GMCs, which would result in a proportional increase of the overlap timescale. By contrast, if the decoupling is caused by feedback-driven GMC destruction, then larger apertures focused on \hii regions would not have an increased probability of containing GMCs, and the overlap timescale would be independent of the aperture size.

\begin{figure*}
\includegraphics[width=\hsize]{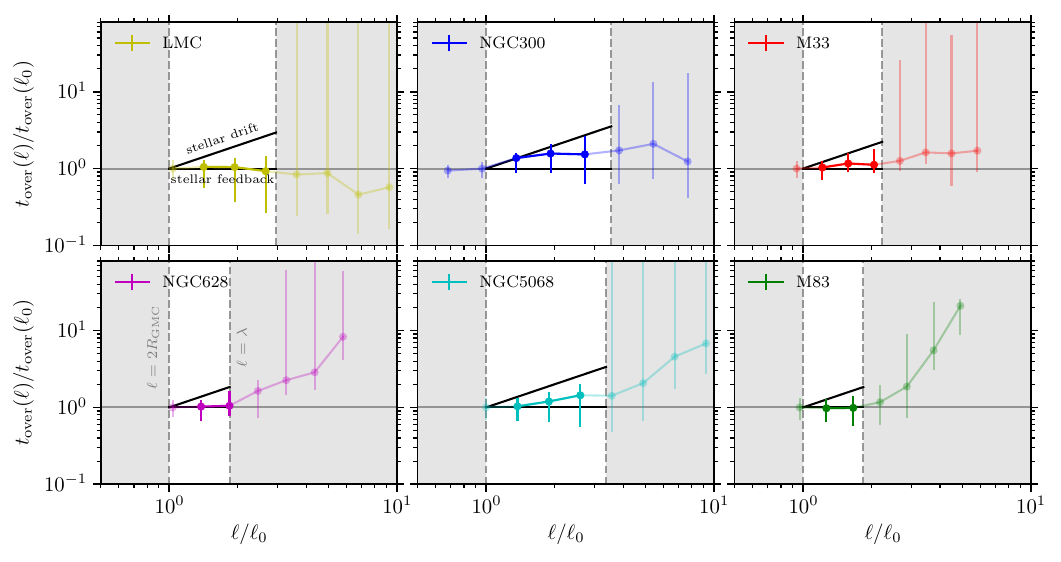}%
\caption{\label{fig:fits}
Relative change of the measured overlap timescale as a function of the spatial resolution, shown for the six galaxies in our sample. The data points with error bars show the measurements. The relevant range of spatial resolutions is bracketed by the two grey-shaded areas -- the transparent data points in the grey-shaded ranges either probe sub-GMC scales (towards the left) or suffer from blending with neighbouring clouds by exceeding the region separation length (towards the right). The small relevant range of resolutions illustrates the technical challenge of carrying out this experiment. Within this range bracketed by the grey-shaded regions, the horizontal solid line indicates the expectation if stellar feedback causes the GMC-\hii region decorrelation, whereas the inclined solid line indicates the expectation if stellar drift drives apart GMCs and their nascent stellar populations. The figure shows that the data are more consistent with stellar feedback-driven GMC dispersal than with stellar drift. This is statistically quantified in \S\ref{sec:stats} and in Table~\ref{tab:stats}.
}
\end{figure*}

The results of our experiment are shown in \autoref{fig:fits}. Per galaxy, only a few ($2{-}3$) data points fall within the range of spatial resolutions where they help distinguish both the feedback and drift scenarios, but across all galaxies a total of 16 data points are suitable for the comparison. Despite the data scarcity, we clearly see that most galaxies have constant overlap timescales of the relevant resolution range, indicating that feedback-driven GMC dispersal is favoured over stellar drift-driven decoupling between GMCs and \hii regions. The overlap timescales only start to increase at resolutions coarser than the region separation length ($\ell>\lambda$), reflecting an artificial increase due to the blending of neighbouring regions (see e.g.\ Appendix B2 of \citealt{kruijssen18}). This behaviour is expected when GMCs are destroyed by stellar feedback.

\subsection{Statistical tests} \label{sec:stats}
We quantify the agreement between the data and the expectations for the feedback and drift hypotheses using a variety of statistical tests. Although we note that these tests are not strictly independent, we present the results of all relevant tests commonly used in hypothesis testing to avoid making an arbitrary (and potentially biasing) selection of one or two of these. These tests are performed in logarithmic space, i.e.\ by using the quantity $\log{[\tover(\ell)/\tover(\ell_0)]}$. Visual inspection of the probability distribution function of $\tover$ reveals close-to-Gaussian distributions. For several of the tests, we therefore make use of $z$-values, which express by how many standard errors the data are separated from the model expectation:
\begin{equation}
    z=\frac{x-\mu}{\sigma} ,
\end{equation}
where $x$ is the observation, $\mu$ is the value expected for a given model (i.e.\ feedback or drift), and $\sigma$ is the $1\sigma$ uncertainty on the observation. When calculating the $z$-values, we account for the asymmetry of the error bars. The results of all statistical tests are presented in Table~\ref{tab:stats}. In the following, we first describe the full set of tests before discussing the results.

\subsubsection{Kolmogorov-Smirnov test}
We first evaluate the null hypotheses that the $z$-values for the feedback and drift scenarios are drawn from a standard normal distribution ${\cal N}(0, 1)$, by employing the two-sample Kolmogorov-Smirnov (KS) test. While it is generally recommended for larger sample sizes and subject to caveats\footnote{See e.g.\ \url{https://asaip.psu.edu/Articles/beware-the-kolmogorov-smirnov-test/}}, the KS test is very commonly used for hypothesis testing due to its simplicity and flexibility. To perform the test, a reference sample of $10^5$ data points is generated from ${\cal N}(0, 1)$, against which the empirical cumulative distribution functions (CDFs) of the $z$-values are compared. The KS statistic quantifies the maximum distance between the empirical and the reference CDFs, and the associated $p$-value expresses the probability of observing such a statistic under the null hypothesis.

\subsubsection{Paired $t$-test}
We continue our analysis with a paired sample $t$-test for the feedback and drift scenarios, which provides a statistical evaluation of the mean differences between the observed data points and the respective expectations for either scenario. The $t$-statistic is calculated as $t = \bar{d}/(s_d/\sqrt{n})$, where $\bar{d}$ is the mean of the paired differences, $s_d$ is their standard deviation, and $n$ is the number of observations. The $t$-statistic effectively represents the number of standard deviations that the sample mean deviates from the null hypothesis that there is no difference between the observations and expectation. The associated $p$-value indicates the probability of observing a $t$-statistic at least as extreme as the one calculated, assuming that the null hypothesis is true.

The $t$-test relies on the standard deviation of the paired differences rather than on the individual error bars of the data points. Therefore, a scenario for which the data exhibit greater variations in $d$ may appear more consistent with the data according to the $t$-test, simply because the data have a larger standard deviation. This counterintuitive property of the $t$-test should be kept in mind when evaluating the results.

\subsubsection{Combined probability test}
We conduct further statistical analysis to assess whether the mean and standard deviation of the $z$-values for both the feedback and drift scenarios significantly deviate from the expected values of 0 and 1, respectively, under the assumption of a standard normal distribution ${\cal N}(0, 1)$. We use a one-sample $t$-test to determine if the mean of the $z$-values differs significantly from 0, resulting in the usual $t$-statistic and $p$-value for each of the scenarios.

Additionally, we apply a $\chi^2$ test to determine if the variance of the $z$-values differs significantly from unity. This is done by first computing the sample variance of the $z$-values and dividing this quantity by the number of degrees of freedom ($n-1$, where $n$ is the sample size). The resulting $\chi^2$ statistic is then compared to the CDF of the $\chi^2$ distribution with $n-1$ degrees of freedom. We calculate the $p$-value by finding the probability that a $\chi^2$ distribution with $n-1$ degrees of freedom would yield a value at least as extreme as the observed $\chi^2$ statistic. This probability is given by the integral of the tail of the $\chi^2$ distribution beyond the observed $\chi^2$ statistic.

To synthesise the results from both the mean and variance tests, we combine both $p$-values using Fisher's method. We first calculate the quantity $-2 \sum\ln(p_i)$, where $p_i$ are the individual $p$-values, and then compare the result to a $\chi^2$ distribution for four degrees of freedom (corresponding to twice the number of tests, which here are constituted by the $t$-test and the $\chi^2$ test). The integral of the tail of the distribution beyond this value then yields the combined $p$-value, which provides a single metric to evaluate the overall deviation of the mean and standard deviation of the $z$-values from a normal distribution.

\subsubsection{Sample z-score}
To evaluate the deviation of the sample means from the expectations for both the feedback and drift scenarios, we calculate the sample $z$-scores. The sample $z$-score is calculated as $z=\bar{z}\sqrt{n}$, where $\bar{z}$ is the mean of the $z$-values. This transformation yields the $z$-score of the sample mean, which indicates how many standard deviations the sample mean is from the null hypothesis mean.

\subsubsection{$\chi^2$ test}
To assess the goodness of fit between the observed data points and the expectations for the feedback and drift scenarios, we employ a $\chi^2$ test. The $\chi^2$ statistic is calculated for each scenario by summing the squared differences between the observed and expected values, normalised by the squared error terms. This calculation incorporates the asymmetric error bars associated with each data point. The resulting $\chi^2$ statistic is then divided by the number of observations $n$. The corresponding $p$-value is calculated as described above, using a $\chi^2$ distribution with $n$ degrees of freedom, which is appropriate here because the expectations for both scenarios are fixed and involve no free parameters. This $p$-value reflects the probability of observing a $\chi^2$ statistic at least as extreme as the one obtained under the null hypothesis that the observed data are consistent with the expectations.

\subsubsection{Bayesian comparison} \label{sec:bayes}

\begin{table*}
\caption{Statistical test results}
\label{tab:stats}
\centering
\begin{tabular}{l|c|cc|cc|cc|cc|cc|cc}
\hline
Galaxy & Number      & \multicolumn{2}{c|}{KS-test $\log p$-value} & \multicolumn{2}{c|}{$t$-test $\log p$-value} & \multicolumn{2}{c|}{$z$-values $\log p$-value} & \multicolumn{2}{c|}{sample $z$-score}  & \multicolumn{2}{c|}{$\chi^2$-test $\log p$-value}  & \multicolumn{2}{c}{Bayes $\log$ post.\ prob.}  \\
       & of data & feedback & drift & feedback & drift & feedback & drift & feedback & drift & feedback & drift & feedback & drift  \\
\hline
LMC & 3 & -0.40 & -2.34 & -0.05 & -1.02 & -0.01 & -1.00 & -0.03 & -3.45 & -0.00 & -2.23 & -0.01 & -1.66 \\
NGC300 & 3 & -1.19 & -0.47 & -1.97 & -0.56 & -1.04 & -0.32 & 1.14 & -1.00 & -0.14 & -0.18 & -0.58 & -0.13 \\
M33 & 3 & -0.60 & -1.82 & -0.98 & -0.95 & -0.42 & -1.22 & 0.65 & -1.82 & -0.04 & -0.47 & -0.09 & -0.73 \\
NGC628 & 2 & -0.32 & -1.49 & -0.69 & -0.72 & -0.21 & -0.73 & 0.16 & -1.90 & -0.01 & -0.79 & -0.09 & -0.73 \\
NGC5068 & 3 & -0.59 & -2.09 & -0.74 & -1.43 & -0.45 & -1.07 & 0.43 & -2.64 & -0.01 & -1.18 & -0.13 & -0.58 \\
M83 & 2 & -0.30 & -1.24 & -0.71 & -0.69 & -0.22 & -0.46 & -0.09 & -1.70 & -0.00 & -0.66 & -0.09 & -0.74 \\
\hline
All & 16 & -2.49 & -6.83 & -2.20 & -5.20 & -1.54 & -5.38 & 0.97 & -5.12 & -0.00 & -1.91 & -0.00 & -3.60 \\
\hline
\end{tabular}
\end{table*}

We conclude the suite of statistical tests by comparing the feedback and drift scenarios in a Bayesian context. This is done by calculating their respective posterior probabilities based on the observed data. We assign prior probabilities $p_{\rm fb} = 0.5$ and $p_{\rm drift} = 0.5$, to reflect a lack of preference for either scenario. The likelihood of each scenario given the data, ${\cal L}_{\rm fb}$ and ${\cal L}_{\rm drift}$, is computed as the product of the probability density functions of the normal distribution, with means corresponding to the expectations and standard deviations by the observed error bars. These likelihoods represent the probability of the observed data under the assumption that a particular scenario is true. The marginal likelihoods for each scenario, ${\cal M}_{\rm fb}$ and ${\cal M}_{\rm drift}$, are then calculated by weighting the likelihoods by the respective prior probabilities (which are both equal to $0.5$ in this case). The total evidence across both scenarios is then defined as $p_{\rm tot}={\cal M}_{\rm fb}+{\cal M}_{\rm drift}$. Finally, the posterior probabilities are obtained as $p_{\rm fb}={\cal M}_{\rm fb}/p_{\rm tot}$ and $p_{\rm drift}={\cal M}_{\rm drift}/p_{\rm tot}$. These posterior probabilities reflect the degree of belief in each scenario given the data and the prior information.

\subsubsection{Test results}

Table~\ref{tab:stats} lists the results of these different tests, showing that we obtain qualitatively consistent results across the full suite of statistics. The small number of data points per galaxy causes the comparison to be inconclusive in some cases, but when performing the tests described above on the full sample of 16 data points, the probabilities differ significantly between the stellar feedback and drift scenarios. Across all tests and galaxies, the range of probability that the data are consistent with stellar feedback-driven GMC dispersal is $p_{\rm fb}=3.2\times10^{-3}{-}1$, with a logarithmic average of $p_{\rm fb}=5.7\times10^{-2}$. By contrast, the range of probability that the data are consistent with the stellar drift-driven decoupling of GMCs and \hii regions is $p_{\rm fb}=1.5\times10^{-7}{-}1.2\times10^{-2}$, with a logarithmic average of $p_{\rm fb}=2.6\times10^{-5}$. Comparing the probability ratios between all tests shows that the small-scale decorrelation between GMCs and \hii regions is approximately 100 to 20,000 times more likely to be driven by stellar feedback than by stellar drift, with the logarithmic average across all tests indicating that it is about 2,000 times more likely to be feedback-driven.

\subsection{Variable drift velocities} \label{sec:alpha}
One of the key assumptions made in this work is that the drift velocity is constant. This need not be the case, as the velocity difference between a GMC and its nascent \hii region might decelerate (e.g.\ by the mutual gravitational attraction) or accelerate (e.g.\ by differential rotation in the galactic potential, or by stellar feedback from the \hii region acting on the GMC). We now test whether the data presented in \autoref{fig:fits} could be reconciled with the drift scenario by adopting a suitable acceleration law.

For this purpose, we assume a simple power law acceleration relation for describing the drift out of an aperture with a diameter $\ell\geq2\rgmc$:
\begin{equation}
\label{eq:vdrift}
v_{\rm drift}(\ell)=v_0\left(\frac{\ell}{\ell_0}\right)^\alpha ,
\end{equation}
where the initial conditions are defined as before, i.e.\ $\ell_0=2\rgmc$ and $t_0=\tover(\ell_0)$. For simplicity, we assume that the velocity at $t<t_0$ is constant, such that $v_0=\ell_0/t_0$. Assuming the acceleration law of equation~(\ref{eq:vdrift}), the expected relative increase of the overlap time with the spatial resolution follows by integration of the equation of motion as
\begin{eqnarray}
\label{eq:alpha}
\left[\frac{\tover(\ell)}{\tover(\ell_0)}\right]_{\rm drift}&=&1+\frac{1}{t_0}\int_{\ell_0}^\ell \frac{{\rm d}\ell' }{v_{\rm drift}(\ell')}  \\
&=&\left\{\begin{array}{ll}
\frac{1}{1-\alpha}\left[\left(\frac{\ell}{\ell_0}\right)^{1-\alpha}-1\right]+1 & \mbox{if } \alpha\neq1 \\
\ln{\left(\frac{\ell}{\ell_0}\right)}+1 & \mbox{if } \alpha=1 .
\end{array} \right.\nonumber
\end{eqnarray}
This is a generalised form of equation~(\ref{eq:toverdrift}), to which it reduces for a constant drift velocity ($\alpha=0$). These expressions allow us to determine which values of $\alpha$ are needed to make the data from \autoref{fig:fits} consistent with the stellar drift scenario.

\begin{figure}
\includegraphics[width=\hsize]{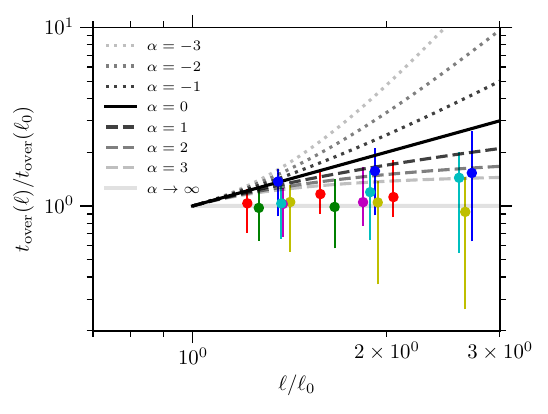}%
\caption{
\label{fig:alpha}
Relative change of the overlap timescale as a function of the spatial resolution, comparing the observational data (coloured points) to the expectations in the stellar drift scenario (lines). The comparison is made for a variety of different acceleration laws, parameterised through $\alpha$ from equation~(\ref{eq:vdrift}) as indicated by the legend. The colours of the data points are the same as in \autoref{fig:fits}, enabling the straightforward identification of each galaxy. Only data points for the relevant range of spatial resolutions ($2\rgmc\leq\ell\leq\lambda$) are shown. The lack of change in $\tover$ towards coarser resolutions ($\ell>\ell_0$) means that making the data consistent with the stellar drift scenario would require a superlinear increase of the drift velocity with distance ($\alpha>1$). There is no known physical agent for such an acceleration.
}
\end{figure}

\autoref{fig:alpha} compares the observational data from \autoref{fig:fits} to the solutions of equation~(\ref{eq:alpha}). The lack of change in $\tover$ towards coarser resolutions ($\ell>\ell_0$) means that making the data consistent with the stellar drift scenario would require high values of $\alpha>1$ (and all galaxies are actually consistent with $\alpha\rightarrow\infty$). This implies that the drift velocity would need to increase superlinearly with distance, and that the acceleration itself would also increase with distance. Differential rotation in the galactic potential (`shear') is known to drive a velocity differential that increases linearly with the radial distance ($\alpha=1$; e.g.\ \citealt{kruijssen19c}), whereas a possible acceleration due to stellar feedback acting on the GMC would decrease with distance ($\alpha<1$), rather than exhibiting the increase with distance ($\alpha>1$) seen here.

\begin{figure}
\includegraphics[width=\hsize]{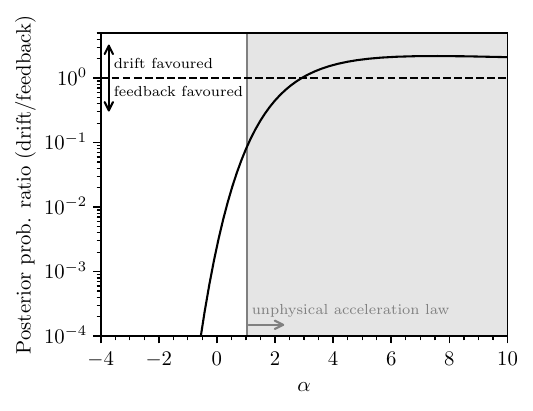}%
\caption{
\label{fig:alpha2}
Posterior probability ratio of the drift and feedback scenarios, visualising when the drift scenario with a variable drift velocity might be favoured over the feedback scenario, shown as a function of the acceleration law slope $\alpha$ (solid line). Equal probability of the feedback and drift scenarios ($p_{\rm drift}/p_{\rm fb}=1$) is indicated by the horizontal dashed line. Unphysical values of $\alpha>1$ are indicated by the shaded region, highlighting that stellar feedback-driven GMC dispersal is favoured ($p_{\rm drift}/p_{\rm fb}<1$) for all physical drift acceleration laws ($\alpha\leq1$). As expected, the posterior probability ratio converges to $1$ when $\alpha\rightarrow\infty$.
}
\end{figure}

The result for a variable drift velocity is further visualised in \autoref{fig:alpha2}, which repeats the Bayesian comparison of \S\ref{sec:bayes} as a function of $\alpha$.\footnote{We symmetrise the error bars in logarithmic space before carrying out the comparison. This is done because the probability would otherwise exhibit discontinuities at values of $\alpha$ for which the lines in \autoref{fig:alpha} precisely match a data point. At such a value, the error bar used in the comparison changes between the upward and downward error, implying that asymmetric error bars would cause a jump in probability.} The stellar drift scenario is disfavoured ($p_{\rm drift}/p_{\rm fb}<1$) for all physical values of $\alpha\leq1$. In summary, using a variable drift velocity to reconcile the observations with the stellar drift scenario would require conditions that are unparalleled in galactic dynamics or stellar feedback physics. This means that stellar drift is ruled out by the observations even if the drift velocity is allowed to vary.

\section{Discussion}
\label{sec:disc}
The reasonable hypothesis put forward by \citetalias{koda23} is that the decorrelation between molecular gas and star formation at $\sim100$-pc scales may originate from stellar drift. The measurements presented in the current work disprove this stellar drift hypothesis at the $3{-}4\sigma$ level and show that the decorrelation is feedback-driven. This conclusion holds even when the drift velocity is allowed to vary because the data would require the stellar drift to follow an unphysical acceleration law. Due to the limited spatial dynamic range over which our measurements can be made, we can draw these conclusions only for one galaxy individually (the LMC). However, when all measurements are combined, the data enable an unambiguous interpretation. Quantitatively, we find that the drift hypothesis is inconsistent with the data, with a likelihood of roughly 1\% (as indicated by a $\chi^2$ test), whereas a Bayesian likelihood comparison between both scenarios shows that the feedback scenario is 4,000 times more likely to explain the data than drift. Across all statistical tests, we find that the small-scale decorrelation between GMCs and \hii regions is on average 2,000 times more likely to be driven by stellar feedback than by stellar drift. The experiment presented here adds the necessary empirical and statistical evidence to earlier discussions on this topic by \citet{kruijssen14} and \citet{chevance20}, who gave physical reasons why stellar drift is likely a negligible contributor to the gas-star formation decorrelation compared to stellar feedback-driven GMC dispersal.

In previous studies, we did not distinguish between GMC destruction by feedback or kinetic dispersal \citep{kruijssen19}. Our new results suggest that GMCs undergo feedback-driven dispersal at the end of their lifetimes to the point that they either dissolve altogether or are dispersed into a diffuse molecular phase. This paper also includes the first use of the uncertainty principle methodology to characterise the GMC lifecycle in the nearby spiral galaxy M83, for which \citet{koda23b} estimated GMC lifetimes $\sim100~\myr$. Our analysis of the decorrelation between GMCs and \hii regions shows that the GMC lifetime is considerably shorter, at $\tgmc=33.2^{+5.6}_{-3.7}~\myr$, and the `feedback timescale' needed for stellar feedback to clear the parent GMC (in this paper referred to as the `overlap timescale', by exception to other papers using this method) is $\tover=3.9^{+1.1}_{-0.6}~\myr$. These numbers are consistent with the range of values found in $50{-}100$ other nearby galaxies \citep[e.g.][]{kruijssen19,chevance20,kim22}, for which the GMC lifetimes range from $5{-}35~\myr$ and the feedback timescales range from $1{-}6~\myr$.

While the main body of their paper presents the stellar drift hypothesis, \citetalias{koda23} conclude by attempting to identify further reasons why the GMC lifetimes measured with the uncertainty principle methodology might possibly be underestimated. Unfortunately, these further reasons are rooted in factual misrepresentations of the uncertainty principle methodology. Because an underestimation of the measured timescales would also call into question the physical conclusions of the experiment presented in this paper, we take the opportunity to briefly correct these misrepresentations here, with references to earlier studies where these potential issues have been dispelled more extensively.
\begin{enumerate}
\item
\textit{The reference timescale $t_{\rm star,ref}$ from \citet{haydon20} would ``[be] estimated under an assumption that all stars in a [region] form simultaneously'', and therefore would cause the duration of the entire GMC lifecycle to be underestimated} \citepalias{koda23}. --- Fundamentally, the uncertainty principle methodology measures the relative timescale ratio between the lifetimes of GMCs and the SFR tracer (usually referring to ionised emission from \hii regions). The emission timescale of the SFR tracer is known from stellar evolution and provides the calibration to which all other timescales are anchored \citep{haydon20_ext,haydon20}. As discussed by \citet[\S3.1; and also in previous observational applications, e.g.\ the Methods section of \citealt{kruijssen19} and \S3.2 of \citealt{chevance20}]{haydon20}, the reference timescale $t_{\rm star,ref}$ refers exclusively to the lifetime of the \hii region after it is devoid of molecular gas and the possibility of further star formation has ceased. Any time during which the GMC may be forming stars over an extended period of time thus precedes the duration of the reference timescale. Because the physical definition of $t_{\rm star,ref}$ specifically refers to the GMC-less lifetime of \hii regions, it enables preceding age spreads of any duration. The total duration of the stellar timescale is defined as $t_{\rm star}=\tover+t_{\rm star,ref}$, where $t_{\rm star,ref}$ is fixed, but there is no limit placed on any age spread encapsulated by $\tover$. Therefore, the concern raised by \citetalias{koda23} is not justified.
\item
\textit{The uncertainty principle methodology ``uses the [gas-to-SFR tracer] flux ratio, and thus is biased toward the population of the brighter clouds in spiral arms'', resulting in short cloud lifetimes compared to ``long-lived'' inter-arm clouds} \citepalias{koda23}. --- While using the gas-to-SFR tracer flux ratio in itself does not lead to a flux bias, it is true that its specific use in the uncertainty principle methodology results in flux-weighted GMC lifetimes. However, it has been used to measure GMC lifetimes across $50{-}100$ galaxies and individual subregions of these, many of which do not contain spiral arms. This motivated \citet{chevance20} to write ``[we] do not find any dependence of the measured evolutionary timelines on the strength or the number of spiral arms in the galaxies of our sample. [...] As a result, we suggest that the offsets between molecular clouds and \hii regions perpendicular to spiral arms that have been used to infer evolutionary timescales are driven primarily by cloud evolution and feedback rather than by dynamical drift''. Galaxies with more pronounced spiral arms (e.g.\ M51 and M83) have longer GMC lifetimes (see \S\ref{sec:results} and \citealt{chevance20}), contrary to the suggestion made by \citetalias{koda23}. Finally, Romanelli et al.\ (in prep.) have applied the uncertainty principle methodology separately to GMCs in spiral arms and those in inter-arm regions, and find no statistically significant difference in GMC lifetimes between both environments. In conclusion, any bias towards GMCs in spiral arms does not cause the GMC lifetime measurements to be underestimated.
\item
\textit{The region separation length $\lambda$ would ``[depend] on the global distribution of the star formation activity'', therefore it ``does not trace the local physical processes of star formation''} \citepalias{koda23}. --- While $\lambda$ is referred to as the `region separation length', it is important to realise that this is measured locally, as it measures the divergence scale of the tuning fork diagram, i.e.\ the size scale at which GMCs and \hii regions decorrelate (see e.g.\ \S3.2.11 and \S7.2.1 of \citealt{kruijssen18}). Therefore, it differs fundamentally from the geometric mean separation length that may be obtained as $l=2\sqrt{A\pi N}$, with $A$ the total area and $N$ the number of GMCs or \hii regions. In galaxies with strong morphological features such as spiral arms or bars, $\lambda$ stays close to the actual separation length in the local environment of the GMCs, whereas $l$ would be inflated by the presence of voids. As a result, the uncertainty principle methodology only assumes approximate statistical isotropy on scales $\leq2\lambda$ (which is $150{-}400~\pc$ for the galaxies in this work), which is much smaller than the size scales of the galactic-morphological features observed in nearby galaxies. Indeed, \citet{kruijssen19} explicitly demonstrate that the nearest-neighbour distances of the emission peaks identified in NGC300 are consistent with the measured region separation lengths (see their Methods section, eq.~(9), and Extended Data Figure 7). As such, the uncertainty principle methodology is designed specifically to extract the local physical processes of star formation, independently of the global distribution of mass or (star formation) activity.
\end{enumerate}

This work provides a further example of how the observed decorrelation of GMCs and \hii regions on sub-kpc scales holds key information to help constrain the physical processes governing the GMC lifecycle. The result that stellar feedback rather than stellar drift marks the end of the GMC lifecycle does not only settle a crucial aspect of this cycle, but also provides key insights into the processes that drive galactic ecosystems. The inference that GMCs have finite lifespans and are destroyed by stellar feedback lends further credence to the emerging paradigm in which galaxies are composed of building blocks undergoing vigorous, feedback-driven lifecycles that collectively drive the baryon cycle and regulate star formation within galaxies. The quantifiable difference in probabilities favouring feedback over drift provides a robust statistical basis for this picture, and the ongoing advances in data quality and statistical methodologies promise to refine these insights further. Such advancements, building on the uncertainty principle methodology (e.g.\ by integrating modern machine learning techniques; Chevance \& Kruijssen in prep.), may provide an increasingly nuanced understanding of the interplay between the GMC lifecycle and galaxy evolution.

\acknowledgments
\noindent\textit{Acknowledgments:}
We thank Jaeyeon Kim for providing the source files from \citet{kim21,kim22}, and Jin Koda, Jonathan Tan, and Mark Krumholz for helpful discussions. J.M.D.K.\ gratefully acknowledges funding from the European Research Council (ERC) under the European Union's Horizon 2020 research and innovation programme via the ERC Starting Grant MUSTANG (grant agreement number 714907).
M.C., L.R.\ and A.R.\ gratefully acknowledge funding from the Deutsche Forschungsgemeinschaft (DFG, German Research Foundation) through an Emmy Noether Research Group (grant number CH2137/1-1).
A.G.\ acknowledges support from the NSF through AST 2008101, AST 220651, and CAREER 2142300.
COOL Research DAO is a Decentralised Autonomous Organisation supporting research in astrophysics aimed at uncovering our cosmic origins.\\

\noindent\textit{Software:} {
\package{matplotlib} \citep{hunter07},
\package{numpy} \citep{vanderwalt11},
\package{pandas} \citep{reback20}
\package{scipy} \citep{jones01}
\package{seaborn} \citep{waskom20}.
}\\

\noindent\textit{Author contributions:}
J.M.D.K.\ carried out the experiment design and wrote the text, to which M.C.\ and S.N.L. contributed. M.C.\ carried out the physical analysis of the data and produced the figures and tables, to which J.M.D.K. and S.N.L.\ contributed. All authors contributed to aspects of the analysis, the interpretation of the results, and the preparation of the manuscript.

\bibliographystyle{aasjournal}
\bibliography{mybib}

\begin{thebibliography}{}
\expandafter\ifx\csname natexlab\endcsname\relax\def\natexlab#1{#1}\fi
\providecommand{\url}[1]{\href{#1}{#1}}
\providecommand{\dodoi}[1]{doi:~\href{http://doi.org/#1}{\nolinkurl{#1}}}
\providecommand{\doeprint}[1]{\href{http://ascl.net/#1}{\nolinkurl{http://ascl.net/#1}}}
\providecommand{\doarXiv}[1]{\href{https://arxiv.org/abs/#1}{\nolinkurl{https://arxiv.org/abs/#1}}}

\bibitem[{{Bigiel} {et~al.}(2008){Bigiel}, {Leroy}, {Walter}, {Brinks}, {de
  Blok}, {Madore}, \& {Thornley}}]{bigiel08}
{Bigiel}, F., {Leroy}, A., {Walter}, F., {et~al.} 2008, \aj, 136, 2846,
  \dodoi{10.1088/0004-6256/136/6/2846}

\bibitem[{{Chevance} {et~al.}(2023){Chevance}, {Krumholz}, {McLeod},
  {Ostriker}, {Rosolowsky}, \& {Sternberg}}]{chevance23}
{Chevance}, M., {Krumholz}, M.~R., {McLeod}, A.~F., {et~al.} 2023, in
  Astronomical Society of the Pacific Conference Series, Vol. 534, Protostars
  and Planets VII, ed. S.~{Inutsuka}, Y.~{Aikawa}, T.~{Muto}, K.~{Tomida}, \&
  M.~{Tamura}, 1, \dodoi{10.48550/arXiv.2203.09570}

\bibitem[{{Chevance} {et~al.}(2020{\natexlab{a}}){Chevance}, {Kruijssen},
  {Hygate}, {Schruba}, {Longmore}, {Groves}, {Henshaw}, {Herrera}, {Hughes},
  {Jeffreson}, {Lang}, {Leroy}, {Meidt}, {Pety}, {Razza}, {Rosolowsky},
  {Schinnerer}, {Bigiel}, {Blanc}, {Emsellem}, {Faesi}, {Glover}, {Haydon},
  {Ho}, {Kreckel}, {Lee}, {Liu}, {Querejeta}, {Saito}, {Sun}, {Usero}, \&
  {Utomo}}]{chevance20}
{Chevance}, M., {Kruijssen}, J.~M.~D., {Hygate}, A. P.~S., {et~al.}
  2020{\natexlab{a}}, \mnras, 493, 2872, \dodoi{10.1093/mnras/stz3525}

\bibitem[{{Chevance} {et~al.}(2020{\natexlab{b}}){Chevance}, {Kruijssen},
  {Vazquez-Semadeni}, {Nakamura}, {Klessen}, {Ballesteros-Paredes}, {Inutsuka},
  {Adamo}, \& {Hennebelle}}]{chevance20b}
{Chevance}, M., {Kruijssen}, J.~M.~D., {Vazquez-Semadeni}, E., {et~al.}
  2020{\natexlab{b}}, \ssr, 216, 50, \dodoi{10.1007/s11214-020-00674-x}

\bibitem[{{Chevance} {et~al.}(2022){Chevance}, {Kruijssen}, {Krumholz},
  {Groves}, {Keller}, {Hughes}, {Glover}, {Henshaw}, {Herrera}, {Kim}, {Leroy},
  {Pety}, {Razza}, {Rosolowsky}, {Schinnerer}, {Schruba}, {Barnes}, {Bigiel},
  {Blanc}, {Dale}, {Emsellem}, {Faesi}, {Grasha}, {Klessen}, {Kreckel}, {Liu},
  {Longmore}, {Meidt}, {Querejeta}, {Saito}, {Sun}, \& {Usero}}]{chevance22}
{Chevance}, M., {Kruijssen}, J.~M.~D., {Krumholz}, M.~R., {et~al.} 2022,
  \mnras, 509, 272, \dodoi{10.1093/mnras/stab2938}

\bibitem[{{Corbelli} {et~al.}(2017){Corbelli}, {Braine}, {Bandiera},
  {Brouillet}, {Combes}, {Druard}, {Gratier}, {Mata}, {Schuster}, {Xilouris},
  \& {Palla}}]{corbelli17}
{Corbelli}, E., {Braine}, J., {Bandiera}, R., {et~al.} 2017, \aap, 601, A146,
  \dodoi{10.1051/0004-6361/201630034}

\bibitem[{{Dobbs} {et~al.}(2015){Dobbs}, {Pringle}, \&
  {Duarte-Cabral}}]{dobbs15}
{Dobbs}, C.~L., {Pringle}, J.~E., \& {Duarte-Cabral}, A. 2015, \mnras, 446,
  3608, \dodoi{10.1093/mnras/stu2319}

\bibitem[{{Elmegreen}(2000)}]{elmegreen00}
{Elmegreen}, B.~G. 2000, \apj, 530, 277, \dodoi{10.1086/308361}

\bibitem[{{Fujimoto} {et~al.}(2019){Fujimoto}, {Chevance}, {Haydon},
  {Krumholz}, \& {Kruijssen}}]{fujimoto19}
{Fujimoto}, Y., {Chevance}, M., {Haydon}, D.~T., {Krumholz}, M.~R., \&
  {Kruijssen}, J.~M.~D. 2019, \mnras, 487, 1717, \dodoi{10.1093/mnras/stz641}

\bibitem[{{Hartmann}(2001)}]{hartmann01}
{Hartmann}, L. 2001, \aj, 121, 1030, \dodoi{10.1086/318770}

\bibitem[{{Haydon} {et~al.}(2020{\natexlab{a}}){Haydon}, {Fujimoto},
  {Chevance}, {Kruijssen}, {Krumholz}, \& {Longmore}}]{haydon20_ext}
{Haydon}, D.~T., {Fujimoto}, Y., {Chevance}, M., {et~al.} 2020{\natexlab{a}},
  \mnras, 497, 5076, \dodoi{10.1093/mnras/staa2162}

\bibitem[{{Haydon} {et~al.}(2020{\natexlab{b}}){Haydon}, {Kruijssen},
  {Chevance}, {Hygate}, {Krumholz}, {Schruba}, \& {Longmore}}]{haydon20}
{Haydon}, D.~T., {Kruijssen}, J.~M.~D., {Chevance}, M., {et~al.}
  2020{\natexlab{b}}, \mnras, 498, 235, \dodoi{10.1093/mnras/staa2430}

\bibitem[{Hunter(2007)}]{hunter07}
Hunter, J.~D. 2007, Computing In Science \& Engineering, 9, 90,
  \dodoi{10.1109/MCSE.2007.55}

\bibitem[{{Hygate} {et~al.}(2019){Hygate}, {Kruijssen}, {Chevance}, {Schruba},
  {Haydon}, \& {Longmore}}]{hygate19}
{Hygate}, A. P.~S., {Kruijssen}, J.~M.~D., {Chevance}, M., {et~al.} 2019,
  \mnras, 488, 2800, \dodoi{10.1093/mnras/stz1779}

\bibitem[{{Jeffreson} {et~al.}(2021){Jeffreson}, {Keller}, {Winter},
  {Chevance}, {Kruijssen}, {Krumholz}, \& {Fujimoto}}]{jeffreson21}
{Jeffreson}, S. M.~R., {Keller}, B.~W., {Winter}, A.~J., {et~al.} 2021, \mnras,
  505, 1678, \dodoi{10.1093/mnras/stab1293}

\bibitem[{Jones {et~al.}(2001)Jones, Oliphant, Peterson, {et~al.}}]{jones01}
Jones, E., Oliphant, T., Peterson, P., {et~al.} 2001, {SciPy}: Open source
  scientific tools for {Python}.
\newblock \url{http://www.scipy.org/}

\bibitem[{{Kawamura} {et~al.}(2009){Kawamura}, {Mizuno}, {Minamidani},
  {Filipovi{\'c}}, {Staveley-Smith}, {Kim}, {Mizuno}, {Onishi}, {Mizuno}, \&
  {Fukui}}]{kawamura09}
{Kawamura}, A., {Mizuno}, Y., {Minamidani}, T., {et~al.} 2009, \apjs, 184, 1,
  \dodoi{10.1088/0067-0049/184/1/1}

\bibitem[{{Keller} {et~al.}(2022){Keller}, {Kruijssen}, \&
  {Chevance}}]{keller22}
{Keller}, B.~W., {Kruijssen}, J.~M.~D., \& {Chevance}, M. 2022, \mnras, 514,
  5355, \dodoi{10.1093/mnras/stac1607}

\bibitem[{{Kennicutt}(1998)}]{kennicutt98}
{Kennicutt}, Robert~C., J. 1998, \apj, 498, 541, \dodoi{10.1086/305588}

\bibitem[{{Kim} {et~al.}(2021){Kim}, {Chevance}, {Kruijssen}, {Schruba},
  {Sandstrom}, {Barnes}, {Bigiel}, {Blanc}, {Cao}, {Dale}, {Faesi}, {Glover},
  {Grasha}, {Groves}, {Herrera}, {Klessen}, {Kreckel}, {Lee}, {Leroy}, {Pety},
  {Querejeta}, {Schinnerer}, {Sun}, {Usero}, {Ward}, \& {Williams}}]{kim21}
{Kim}, J., {Chevance}, M., {Kruijssen}, J.~M.~D., {et~al.} 2021, \mnras, 504,
  487, \dodoi{10.1093/mnras/stab878}

\bibitem[{{Kim} {et~al.}(2022){Kim}, {Chevance}, {Kruijssen}, {Leroy},
  {Schruba}, {Barnes}, {Bigiel}, {Blanc}, {Cao}, {Congiu}, {Dale}, {Faesi},
  {Glover}, {Grasha}, {Groves}, {Hughes}, {Klessen}, {Kreckel}, {McElroy},
  {Pan}, {Pety}, {Querejeta}, {Razza}, {Rosolowsky}, {Saito}, {Schinnerer},
  {Sun}, {Tomi{\v{c}}i{\'c}}, {Usero}, \& {Williams}}]{kim22}
---. 2022, \mnras, 516, 3006, \dodoi{10.1093/mnras/stac2339}

\bibitem[{{Kim} {et~al.}(2023){Kim}, {Chevance}, {Kruijssen}, {Barnes},
  {Bigiel}, {Blanc}, {Boquien}, {Cao}, {Congiu}, {Dale}, {Egorov}, {Faesi},
  {Glover}, {Grasha}, {Groves}, {Hassani}, {Hughes}, {Klessen}, {Kreckel},
  {Larson}, {Lee}, {Leroy}, {Liu}, {Longmore}, {Meidt}, {Pan}, {Pety},
  {Querejeta}, {Rosolowsky}, {Saito}, {Sandstrom}, {Schinnerer}, {Smith},
  {Usero}, {Watkins}, \& {Williams}}]{kim23}
---. 2023, \apjl, 944, L20, \dodoi{10.3847/2041-8213/aca90a}

\bibitem[{{Koda} \& {Tan}(2023)}]{koda23}
{Koda}, J., \& {Tan}, J. 2023, arXiv e-prints, arXiv:2308.11717,
  \dodoi{10.48550/arXiv.2308.11717}

\bibitem[{{Koda} {et~al.}(2009){Koda}, {Scoville}, {Sawada}, {La Vigne},
  {Vogel}, {Potts}, {Carpenter}, {Corder}, {Wright}, {White}, {Zauderer},
  {Patience}, {Sargent}, {Bock}, {Hawkins}, {Hodges}, {Kemball}, {Lamb},
  {Plambeck}, {Pound}, {Scott}, {Teuben}, \& {Woody}}]{koda09}
{Koda}, J., {Scoville}, N., {Sawada}, T., {et~al.} 2009, \apjl, 700, L132,
  \dodoi{10.1088/0004-637X/700/2/L132}

\bibitem[{{Koda} {et~al.}(2023){Koda}, {Hirota}, {Egusa}, {Sakamoto}, {Sawada},
  {Heyer}, {Baba}, {Boissier}, {Calzetti}, {Meyer}, {Elmegreen}, {de Paz},
  {Harada}, {Ho}, {Kobayashi}, {Kuno}, {Lee}, {Madore}, {Maeda}, {Mart{\'\i}n},
  {Muraoka}, {Nakanishi}, {Onodera}, {Pineda}, {Scoville}, \&
  {Watanabe}}]{koda23b}
{Koda}, J., {Hirota}, A., {Egusa}, F., {et~al.} 2023, \apj, 949, 108,
  \dodoi{10.3847/1538-4357/acc65e}

\bibitem[{{Kreckel} {et~al.}(2020){Kreckel}, {Ho}, {Blanc}, {Glover}, {Groves},
  {Rosolowsky}, {Bigiel}, {Boqu{\'\i}en}, {Chevance}, {Dale}, {Deger},
  {Emsellem}, {Grasha}, {Kim}, {Klessen}, {Kruijssen}, {Lee}, {Leroy}, {Liu},
  {McElroy}, {Meidt}, {Pessa}, {Sanchez-Blazquez}, {Sandstrom}, {Santoro},
  {Scheuermann}, {Schinnerer}, {Schruba}, {Utomo}, {Watkins}, \&
  {Williams}}]{kreckel20}
{Kreckel}, K., {Ho}, I.~T., {Blanc}, G.~A., {et~al.} 2020, \mnras, 499, 193,
  \dodoi{10.1093/mnras/staa2743}

\bibitem[{{Kruijssen} {et~al.}(2015){Kruijssen}, {Dale}, \&
  {Longmore}}]{kruijssen15}
{Kruijssen}, J.~M.~D., {Dale}, J.~E., \& {Longmore}, S.~N. 2015, \mnras, 447,
  1059, \dodoi{10.1093/mnras/stu2526}

\bibitem[{{Kruijssen} \& {Longmore}(2014)}]{kruijssen14}
{Kruijssen}, J.~M.~D., \& {Longmore}, S.~N. 2014, \mnras, 439, 3239,
  \dodoi{10.1093/mnras/stu098}

\bibitem[{{Kruijssen} {et~al.}(2018){Kruijssen}, {Schruba}, {Hygate}, {Hu},
  {Haydon}, \& {Longmore}}]{kruijssen18}
{Kruijssen}, J.~M.~D., {Schruba}, A., {Hygate}, A. P.~S., {et~al.} 2018,
  \mnras, 479, 1866, \dodoi{10.1093/mnras/sty1128}

\bibitem[{{Kruijssen} {et~al.}(2019{\natexlab{a}}){Kruijssen}, {Schruba},
  {Chevance}, {Longmore}, {Hygate}, {Haydon}, {McLeod}, {Dalcanton}, {Tacconi},
  \& {van Dishoeck}}]{kruijssen19}
{Kruijssen}, J.~M.~D., {Schruba}, A., {Chevance}, M., {et~al.}
  2019{\natexlab{a}}, \nat, 569, 519, \dodoi{10.1038/s41586-019-1194-3}

\bibitem[{{Kruijssen} {et~al.}(2019{\natexlab{b}}){Kruijssen}, {Dale},
  {Longmore}, {Walker}, {Henshaw}, {Jeffreson}, {Petkova}, {Ginsburg},
  {Barnes}, {Battersby}, {Immer}, {Jackson}, {Keto}, {Krieger}, {Mills},
  {S{\'a}nchez-Monge}, {Schmiedeke}, {Suri}, \& {Zhang}}]{kruijssen19c}
{Kruijssen}, J.~M.~D., {Dale}, J.~E., {Longmore}, S.~N., {et~al.}
  2019{\natexlab{b}}, \mnras, 484, 5734, \dodoi{10.1093/mnras/stz381}

\bibitem[{{Leisawitz} {et~al.}(1989){Leisawitz}, {Bash}, \&
  {Thaddeus}}]{leisawitz89}
{Leisawitz}, D., {Bash}, F.~N., \& {Thaddeus}, P. 1989, \apjs, 70, 731,
  \dodoi{10.1086/191357}

\bibitem[{{Leroy} {et~al.}(2008){Leroy}, {Walter}, {Brinks}, {Bigiel}, {de
  Blok}, {Madore}, \& {Thornley}}]{leroy08}
{Leroy}, A.~K., {Walter}, F., {Brinks}, E., {et~al.} 2008, \aj, 136, 2782,
  \dodoi{10.1088/0004-6256/136/6/2782}

\bibitem[{{Leroy} {et~al.}(2013){Leroy}, {Walter}, {Sandstrom}, {Schruba},
  {Munoz-Mateos}, {Bigiel}, {Bolatto}, {Brinks}, {de Blok}, {Meidt}, {Rix},
  {Rosolowsky}, {Schinnerer}, {Schuster}, \& {Usero}}]{leroy13}
{Leroy}, A.~K., {Walter}, F., {Sandstrom}, K., {et~al.} 2013, \aj, 146, 19,
  \dodoi{10.1088/0004-6256/146/2/19}

\bibitem[{{Leroy} {et~al.}(2021{\natexlab{a}}){Leroy}, {Schinnerer}, {Hughes},
  {Rosolowsky}, {Pety}, {Schruba}, {Usero}, {Blanc}, {Chevance}, {Emsellem},
  {Faesi}, {Herrera}, {Liu}, {Meidt}, {Querejeta}, {Saito}, {Sandstrom}, {Sun},
  {Williams}, {Anand}, {Barnes}, {Behrens}, {Belfiore}, {Benincasa},
  {Be{\v{s}}li{\'c}}, {Bigiel}, {Bolatto}, {den Brok}, {Cao}, {Chandar},
  {Chastenet}, {Chiang}, {Congiu}, {Dale}, {Deger}, {Eibensteiner}, {Egorov},
  {Garc{\'\i}a-Rodr{\'\i}guez}, {Glover}, {Grasha}, {Henshaw}, {Ho}, {Kepley},
  {Kim}, {Klessen}, {Kreckel}, {Koch}, {Kruijssen}, {Larson}, {Lee}, {Lopez},
  {Machado}, {Mayker}, {McElroy}, {Murphy}, {Ostriker}, {Pan}, {Pessa},
  {Puschnig}, {Razza}, {S{\'a}nchez-Bl{\'a}zquez}, {Santoro}, {Sardone},
  {Scheuermann}, {Sliwa}, {Sormani}, {Stuber}, {Thilker}, {Turner}, {Utomo},
  {Watkins}, \& {Whitmore}}]{leroy21_survey}
{Leroy}, A.~K., {Schinnerer}, E., {Hughes}, A., {et~al.} 2021{\natexlab{a}},
  \apjs, 257, 43, \dodoi{10.3847/1538-4365/ac17f3}

\bibitem[{{Leroy} {et~al.}(2021{\natexlab{b}}){Leroy}, {Hughes}, {Liu}, {Pety},
  {Rosolowsky}, {Saito}, {Schinnerer}, {Schruba}, {Usero}, {Faesi}, {Herrera},
  {Chevance}, {Hygate}, {Kepley}, {Koch}, {Querejeta}, {Sliwa}, {Will},
  {Wilson}, {Anand}, {Barnes}, {Belfiore}, {Be{\v{s}}li{\'c}}, {Bigiel},
  {Blanc}, {Bolatto}, {Boquien}, {Cao}, {Chandar}, {Chastenet}, {Chiang},
  {Congiu}, {Dale}, {Deger}, {den Brok}, {Eibensteiner}, {Emsellem},
  {Garc{\'\i}a-Rodr{\'\i}guez}, {Glover}, {Grasha}, {Groves}, {Henshaw},
  {Jim{\'e}nez Donaire}, {Kim}, {Klessen}, {Kreckel}, {Kruijssen}, {Larson},
  {Lee}, {Mayker}, {McElroy}, {Meidt}, {Mok}, {Pan}, {Puschnig}, {Razza},
  {S{\'a}nchez-Bl'azquez}, {Sandstrom}, {Santoro}, {Sardone}, {Scheuermann},
  {Sun}, {Thilker}, {Turner}, {Ubeda}, {Utomo}, {Watkins}, \&
  {Williams}}]{leroy21_pipe}
{Leroy}, A.~K., {Hughes}, A., {Liu}, D., {et~al.} 2021{\natexlab{b}}, \apjs,
  255, 19, \dodoi{10.3847/1538-4365/abec80}

\bibitem[{{Lu} {et~al.}(2022){Lu}, {Boyce}, {Haggard}, {Bureau}, {Liang},
  {Liu}, {Choi}, {Cappellari}, {Chemin}, {Chevance}, {Davis}, {Drissen},
  {Elford}, {Gensior}, {Kruijssen}, {Martin}, {Mass{\'e}}, {Robert}, {Ruffa},
  {Rousseau-Nepton}, {Sarzi}, {Savard}, \& {Williams}}]{lu22}
{Lu}, A., {Boyce}, H., {Haggard}, D., {et~al.} 2022, \mnras, 514, 5035,
  \dodoi{10.1093/mnras/stac1583}

\bibitem[{{McKee} \& {Ostriker}(2007)}]{mckee07}
{McKee}, C.~F., \& {Ostriker}, E.~C. 2007, \araa, 45, 565,
  \dodoi{10.1146/annurev.astro.45.051806.110602}

\bibitem[{{McLeod} {et~al.}(2019){McLeod}, {Dale}, {Evans}, {Ginsburg},
  {Kruijssen}, {Pellegrini}, {Ramsay}, \& {Testi}}]{mcleod19}
{McLeod}, A.~F., {Dale}, J.~E., {Evans}, C.~J., {et~al.} 2019, \mnras, 486,
  5263, \dodoi{10.1093/mnras/sty2696}

\bibitem[{{McLeod} {et~al.}(2020){McLeod}, {Kruijssen}, {Weisz}, {Zeidler},
  {Schruba}, {Dalcanton}, {Longmore}, {Chevance}, {Faesi}, \&
  {Byler}}]{mcleod20}
{McLeod}, A.~F., {Kruijssen}, J.~M.~D., {Weisz}, D.~R., {et~al.} 2020, \apj,
  891, 25, \dodoi{10.3847/1538-4357/ab6d63}

\bibitem[{{McLeod} {et~al.}(2021){McLeod}, {Ali}, {Chevance}, {Della Bruna},
  {Schruba}, {Stevance}, {Adamo}, {Kruijssen}, {Longmore}, {Weisz}, \&
  {Zeidler}}]{mcleod21}
{McLeod}, A.~F., {Ali}, A.~A., {Chevance}, M., {et~al.} 2021, \mnras, 508,
  5425, \dodoi{10.1093/mnras/stab2726}

\bibitem[{{Meidt} {et~al.}(2015){Meidt}, {Hughes}, {Dobbs}, {Pety}, {Thompson},
  {Garc{\'\i}a-Burillo}, {Leroy}, {Schinnerer}, {Colombo}, {Querejeta},
  {Kramer}, {Schuster}, \& {Dumas}}]{meidt15}
{Meidt}, S.~E., {Hughes}, A., {Dobbs}, C.~L., {et~al.} 2015, \apj, 806, 72,
  \dodoi{10.1088/0004-637X/806/1/72}

\bibitem[{{Meurer} {et~al.}(2006){Meurer}, {Hanish}, {Ferguson}, {Knezek},
  {Kilborn}, {Putman}, {Smith}, {Koribalski}, {Meyer}, {Oey}, {Ryan-Weber},
  {Zwaan}, {Heckman}, {Kennicutt}, {Lee}, {Webster}, {Bland-Hawthorn},
  {Dopita}, {Freeman}, {Doyle}, {Drinkwater}, {Staveley-Smith}, \&
  {Werk}}]{meurer06}
{Meurer}, G.~R., {Hanish}, D.~J., {Ferguson}, H.~C., {et~al.} 2006, \apjs, 165,
  307, \dodoi{10.1086/504685}

\bibitem[{{Onodera} {et~al.}(2010){Onodera}, {Kuno}, {Tosaki}, {Kohno},
  {Nakanishi}, {Sawada}, {Muraoka}, {Komugi}, {Miura}, {Kaneko}, {Hirota}, \&
  {Kawabe}}]{onodera10}
{Onodera}, S., {Kuno}, N., {Tosaki}, T., {et~al.} 2010, \apjl, 722, L127,
  \dodoi{10.1088/2041-8205/722/2/L127}

\bibitem[{Reback {et~al.}(2020)Reback, McKinney, jbrockmendel, den Bossche,
  Augspurger, Cloud, gfyoung, Sinhrks, Klein, Roeschke, Hawkins, Tratner, She,
  Ayd, Petersen, Garcia, Schendel, Hayden, MomIsBestFriend, Jancauskas,
  Battiston, Seabold, chris b1, h~vetinari, Hoyer, Overmeire, alimcmaster1,
  Dong, Whelan, \& Mehyar}]{reback20}
Reback, J., McKinney, W., jbrockmendel, {et~al.} 2020, pandas-dev/pandas:
  Pandas 1.0.3, v1.0.3,  Zenodo, \dodoi{10.5281/zenodo.3715232}

\bibitem[{{Schruba} {et~al.}(2010){Schruba}, {Leroy}, {Walter}, {Sandstrom}, \&
  {Rosolowsky}}]{schruba10}
{Schruba}, A., {Leroy}, A.~K., {Walter}, F., {Sandstrom}, K., \& {Rosolowsky},
  E. 2010, \apj, 722, 1699, \dodoi{10.1088/0004-637X/722/2/1699}

\bibitem[{{Scoville} \& {Hersh}(1979)}]{scoville79}
{Scoville}, N.~Z., \& {Hersh}, K. 1979, \apj, 229, 578, \dodoi{10.1086/156991}

\bibitem[{{Semenov} {et~al.}(2021){Semenov}, {Kravtsov}, \&
  {Gnedin}}]{semenov21}
{Semenov}, V.~A., {Kravtsov}, A.~V., \& {Gnedin}, N.~Y. 2021, \apj, 918, 13,
  \dodoi{10.3847/1538-4357/ac0a77}

\bibitem[{van~der Walt {et~al.}(2011)van~der Walt, Colbert, \&
  Varoquaux}]{vanderwalt11}
van~der Walt, S., Colbert, S.~C., \& Varoquaux, G. 2011, Computing in Science
  and Engg., 13, 22, \dodoi{10.1109/MCSE.2011.37}

\bibitem[{{Ward} {et~al.}(2020){Ward}, {Chevance}, {Kruijssen}, {Hygate},
  {Schruba}, \& {Longmore}}]{ward20}
{Ward}, J.~L., {Chevance}, M., {Kruijssen}, J.~M.~D., {et~al.} 2020, \mnras,
  497, 2286, \dodoi{10.1093/mnras/staa1977}

\bibitem[{{Ward} {et~al.}(2022){Ward}, {Kruijssen}, {Chevance}, {Kim}, \&
  {Longmore}}]{ward22}
{Ward}, J.~L., {Kruijssen}, J.~M.~D., {Chevance}, M., {Kim}, J., \& {Longmore},
  S.~N. 2022, \mnras, 516, 4025, \dodoi{10.1093/mnras/stac2467}

\bibitem[{Waskom {et~al.}(2020)Waskom, Botvinnik, Ostblom, Lukauskas, Hobson,
  MaozGelbart, Gemperline, Augspurger, Halchenko, Cole, Warmenhoven, de~Ruiter,
  Pye, Hoyer, Vanderplas, Villalba, Kunter, Quintero, Bachant, Martin, Meyer,
  Swain, Miles, Brunner, O'Kane, Yarkoni, Williams, \& Evans}]{waskom20}
Waskom, M., Botvinnik, O., Ostblom, J., {et~al.} 2020, mwaskom/seaborn: v0.10.0
  (January 2020),  Zenodo, \dodoi{10.5281/zenodo.3629446}

\bibitem[{{Zabel} {et~al.}(2020){Zabel}, {Davis}, {Sarzi}, {Nedelchev},
  {Chevance}, {Kruijssen}, {Iodice}, {Baes}, {Bendo}, {Corsini}, {De Looze},
  {de Zeeuw}, {Gadotti}, {Grossi}, {Peletier}, {Pinna}, {Serra}, {van de
  Voort}, {Venhola}, {Viaene}, \& {Vlahakis}}]{zabel20}
{Zabel}, N., {Davis}, T.~A., {Sarzi}, M., {et~al.} 2020, \mnras, 496, 2155,
  \dodoi{10.1093/mnras/staa1513}

\end{thebibliography}

{\onecolumngrid}

\end{document}